\documentstyle[sprocl]{article}

\input{psfig.sty}

\def\Journal#1#2#3#4{{#1} {\bf #2}, #3 (#4)}

\def\NPA{{\em Nucl. Phys.} A}

\def\PRD{{\em Phys. Rev.} D}
\def\ZPC{{\em Z. Phys.} C}

\def\ra{\rightarrow}

\def\be{\begin{equation}}
\def\ee{\end{equation}}
\def\bea{\begin{eqnarray}}
\def\eea{\end{eqnarray}}

\begin{document}

\title{QCD-BAG MASS SPECTRUM AND PHASE TRANSITIONS}
\author{Ahmed TOUNSI, Jean LETESSIER}
\address{Laboratoire de Physique Th\'eorique et Hautes
 Energies\\ Universit\'e Paris 7, Tour 24, 5\`eme \'etage\\ 
2 Place Jussieu, F-75251 Paris Cedex 05, France\\E-mail: tounsi@lpthe.jussieu.fr}
\author{Johann RAFELSKI\footnote{Supported in part
by the U.S. Department of Energy, grant  DE-FG03-95ER40937.}
}
\address{Department of Physics, University of Arizona\\
Tucson, AZ 85721, USA}
\maketitle
\abstracts{We obtain the hadronic mass spectrum in the `bag of bags' statistical 
bootstrap model (BBSBM), implementing the colorless state condition, aside of
baryon and strangeness conservation, using group projection method. We study 
the partition function, investigate the properties of dense hadronic 
matter, and determine the conditions under which the system undergoes 
a phase transition to a deconfined quark-gluon plasma. We show that a 
phase transition cannot occur in the  $N=1$ (Abelian) limit of our model,
and is first order for QCD-like case $N=3$.}

\begin{center}
{\it Dedicated to the memory of Peter A. Carruthers}
\end{center}

\section{Introduction}
Considerable theoretical and experimental effort is being devoted to the  
study of the phase transition between a hadron gas and a quark-gluon plasma (QGP). 
This phase transition is believed to have occurred in the early Universe, and 
is searched for in high energy nuclear collisions in  laboratory.
The fundamental theoretical treatment of the hadronic phase transition relies on 
lattice QCD approximation. This approach exhibits  a phase transition and gives 
the equation of states of both phases. The case of nonzero baryon 
density (finite chemical potential) comprises 
 not yet solved technical  difficulties~\cite{tar}. 

In a phenomenological study of this case one 
starts from the low temperature phase, i.e., 
the hadron phase, and implement the high temperature phase
in a more or less {\em ad hoc} way. This was the case of
the statistical bootstrap model~\cite{hag1} (SBM) when 
a quark component is added to it beyond the critical
curve limiting the existence domain of the hadron phase
~\cite{hag2}. Here we show that it is possible  to extend SBM in
a natural way to encapsulate the high temperature deconfined 
phase. We next recall the SBM  and 
in particular illustrate derivation of the
hadronic mass spectrum.  We introduce the 
modifications leading to a SBM-type model which describe 
both hadron phase and QGP phase in section \ref{BBSBM}, and we study the 
phases and properties of the system at finite baryochemical 
potential in section \ref{PT}.
 
\section{Statistical bootstrap }\label{SBM}
\subsection{The hadronic mass spectrum}\label{subsec:hms}
 The SBM approach can be characterized in three
statements: 1) hadrons are made of hadrons;
 2) compound hadrons and `elementary' hadrons have to be
treated on the same footing: this is the bootstrap;
 3) the main effect of strong interactions is hadron
 production. Hadron-hadron interactions in thermally and 
chemically equilibrated system can be accounted for (nearly) 
completely by considering the abundances of particles present, 
i.e., a strongly interacting
hadron gas can be described as a mixture of infinitely many
ideal gases, weighted by the  hadronic mass spectrum.
The dynamical problem in SBM is the theoretical determination of  
the hadronic mass spectrum,
 i.e., the number of hadron states per unit interval mass.

For pointlike particles in a box of volume $V$ the bootstrap equation for this 
density reads~\cite{fra}
\bea
\rho_{\rm out}(m)&\hspace{-6cm}=&\hspace{-6cm}\sum_{n=2}^{\infty}
 \left({V\over {(2\pi)}^3}\right)^{n-1}
{1\over n!}\prod^n_{i=1}\int\rho_
{\rm in}(m_i)\,dm_i \nonumber\\ &\hspace{0.5cm}\times\displaystyle\int \delta
\bigg (\displaystyle\sum^n_{i=1}~E_i-m\bigg )\delta^3
\bigg(\sum^n_{i=1}\vec{p}{_i}\bigg) d^3p_i\,.
\label{eq:ro}
\eea
In this equation, $\rho_{\rm out}$ represents the total density
of states in the box, while $\rho_
{\rm in}(m_i)$ counts the number of mass states of particle $i$. 
Note that the density of levels is written with center of mass at rest because 
this is the density to be identified with the number of hadron states per unit 
interval of rest mass.

One has to solve Eq.~(\ref{eq:ro}) with the bootstrap condition:
\be
\rho_{out}(m)=\rho_{in}(m)\equiv \rho(m)\,.
\label{bootc}  
\ee
The solution is of the form 
\be
\rho(m)\sim cm^a e^{m/T_0},\qquad  \mbox{with}\qquad  a=-3\,.
\label{eq:sol}
\ee

\subsection{The partition function}\label{subsec:partstb}
The partition function in the Boltzmann approximation can be written
\be
\ln\,Z(V,T)=\frac{VT}{2{\pi}^2}\,\displaystyle\int_0^{\infty}\,
\rho(m)\,m^2\,K_2\left(\frac{m}{T}\right)\,dm\,,
\label{partbol}
\ee
where  $K_2$ is the modified Hankel function.
With a mass spectrum of the form~(\ref{eq:sol}),
one finds for $T\to T_0$,
\bea
\ln\,Z(V,T\to T_0)\approx cV\left(\frac{T_0}{2\pi}\right)^{\frac{3}{2}}
{\left(\frac{T_0-T}{T_0T}\right)}^{-(a+\frac{5}{2})}
\Gamma\left(a+\frac{5}{2},\, ...\right)\,,
\label{zt0}
\eea
and the energy  density,
\be
\varepsilon(T)=\frac{T^2}{V}\frac{\partial\ln\,Z(V,T)}{\partial T}\,,
\label{ed}
\ee
behaves for $T\ra T_0$ as
\be
\varepsilon\approx \left\{ \begin{array}{ll}
      \left(\frac{T_0-T}{T_0T}\right)^{-(a+7/2)} & \mbox{for} \quad a> -7/2\,,\\
      \ln \left(\frac{T_0-T}{T_0T}\right)       & \mbox{for} \quad a=-7/2\,,\\
      \mbox{Const.}       & \mbox{for} \quad a<-7/2\,.\\
      \end{array}\right.
\label{edbeh}
\ee
One sees that for $a\geq -7/2$, $T_0$ is  attained with {\em infinite} 
energy density and appears as a maximal {\em limiting} temperature. 
All energy delivered is converted into particle multiplicity rather than thermal motion.
But for $a < -7/2$, $T_0$ is reached at 
{\em finite}  energy density and the system may undergo a phase transition.
In the original SBM $a >-7/2$ and thus $T_0$  is a limiting temperature~\cite{hag1,hag}.
Interestingly, the hadronic partition function in SBM is singular (a square-root singularity) even
for {\em finite} volume. This last feature can be traced back to the assumption of pointlike hadrons. 
However when the individual hadron volumes are taken into account through the so called 
`excluded volume' correction $T_0$ is  reached at finite energy density but not in the 
usual thermodynamic limit~\cite{hag2}. When baryon number is considered, there is  
not a single transition point $T_0$, but a transition curve 
$T(\mu)$ in the plane $T-\mu$, where $\mu$ is the baryonic 
chemical potential, and $T_0= T(\mu=0)$. Beyond this critical curve the SBM partition 
function becomes complex and extraneous assumptions are needed to implement the new QGP phase. 
This results in a two component model, one component describing the hadron side and another 
the quark-gluon side. However, a single closed analytic model unifying both phases can be built,
provided that we introduce from the start the quark structure of hadrons as extended objects. 

\section{Bag of bags SBM}\label{BBSBM}
\subsection{The bag grand canonical partition function}\label{subsec:bpf}
Hadrons  are considered now as bags of quarks and gluons occupying finite volumes. 
They have to be colorless objects, i.e., $SU_c(3)$ (or $SU_c(N)$) singlets. 
This is an expression of color confinement. To be realistic, we also implement 
baryon number and strangeness conservation effects.
With these constraints, we will determine 
the hadronic mass spectrum in such a bag-of-bags-SBM (BBSBM).
Consider a colorless bag of volume $v$ filled with quarks, anti-quarks and gluons at 
sufficiently high temperature to neglect their interactions (asymptotic freedom). 
The density of states of this bag can be determined by taking the inverse Laplace 
transform of its grand canonical partition function (GCPF). But quarks and gluons 
transform respectively under the fundamental and the adjoint representations of $SU_c(3)$. 
In the GCPF one has to retain only the singlet states. Furthermore, 
as we have noted in Sec.~\ref{subsec:hms}, the density
of levels is defined with the center of mass at rest. 

With the preceding constraints the GCPF, can be derived, using projection technique 
\cite{red}, from the function
\bea\label{eq:gpf}
\widetilde{ Z}_Q(\beta,v,\mu_q, \mu_S) = \displaystyle\int_{SU_{\rm
c}(3)}&&\hspace{-0.7cm}d{\mu(g)} 
\int_v \frac{d^3 R}{v}\left\{Tr_G\,{U_G}(g) e^{-\beta H_G +i \vec{
P_G}\cdot\vec{ R}\phantom{\widehat N_q}}\right\}\\ & &\hspace{-2.5cm}\times
\left\{Tr_q\, {U_q}(g) e^{-\beta H_q +i \vec{ P_q }\cdot\vec{ R} -
i\mu_q {\widehat N_q} }\right\} 
 \left\{Tr_{\bar q}\,{U_{\bar q}}(g) e^{-\beta H_{\bar q} +i \vec{
P_{\bar q}} \cdot\vec{ R} +i\mu_q {\widehat N}_{\bar q} }
\right\} \nonumber\\ 
 & &\hspace{-2.5cm}\times\left\{Tr_s {U_s}(g) e^{-\beta H_s +i \vec{ P_s
}\cdot\vec{ R} -i\mu_s {\widehat N_s} }\right\}
 \left\{Tr_{\bar s}\,{U_{\bar s}}(g) e^{-\beta H_{\bar s} +i \vec{
P_{\bar s}} \cdot\vec{ R} +i\mu_s{\widehat N}_{\bar s} }\right\},
\nonumber 
\eea
where $G$ stands for gluons, $q$ for the quarks $u$ and $d$ and $s$ for the strange quark. 
$\widehat N_q$ and $\widehat N_s$ are the  quark number operators and
$\mu_q$ and $\mu_s$ the corresponding chemical potentials.
$\vec{P_G}$ , $\vec{P_q}$ and $\vec{R}$ are the momenta
of gluons and quarks and their space position inside the bag.
$U(g)$ is a unitary representation of $SU_{\rm c}(3)$ in the Hilbert space
of states~\cite{red}, $g$ being an element of $SU_{\rm c}(3)$. 
The physical grand partition function ${Z}_Q(\beta,v,\mu_q, \mu_s)$ is obtained from 
$\widetilde{ Z}_Q(\beta,v,\mu_q, \mu_s)$ by the so-called `Wick rotation'
$\mu$ $\to -i\beta\mu$.
In spite of its apparent complexity Eq.~(\ref{eq:gpf})
corresponds simply to the product of five partitions functions (five traces) for non 
interacting five particle species ($G$, $q$, $\bar q$, $s$ and $\bar s$). 
The integration over the $SU_c(3)$ group with the Haar measure $d{\mu(g)}$ selects
the $SU_c(3)$ singlet  states and the integration over $R$
in the volume $v$ selects the bag states  of zero momentum.

$\widetilde{ Z}_Q(\beta,v,\mu_q, \mu_s)$ can be written in form ~\cite{tou},
\be
\widetilde{ Z}_{QG}(\beta,v,\mu_q, \mu_s)=\int_{SU_{\rm c}(3)}d{\mu(g)} 
\int_{v/\beta^3} \frac{ d^3 r}{v/\beta^3} e^{\Theta}\,,
\label{eq:ztil}
\ee
where $\vec{r}=\vec{R}/\beta$ and
\be
\Theta=\frac{v}{\pi^2\beta^3}\frac{1}{(1+r^2)^2}
\big [\, d_G\, {\cal G}(g)+d_q{\cal Q}(g,\mu_q)+
d_s{\cal Q}(g,\mu_s)\big]\,,
\label{eq:Theta}
\ee
with
\be
{\cal G}(g)=\sum_{k=0}^{\infty}\frac{1}{k^4}\chi_{1,1} (g^k)\,,
\label{eq:gluons}
\ee
for  gluon contribution and
\be
{\cal Q}(g,\mu)=2\Re \sum_{k=0}^{\infty}\frac{(-1)^{k-1
}}{k^4}e^{ik\mu}\chi_{(1,0)}(g^k)\,,
\label{eq:quarks}
\ee
for  quark contribution.  $\chi_{1,0}(g)$ and 
$\chi_{1,1}(g)=|\chi_{1,0}(g)|^2 -1$
are respectively the characters of the fundamental and
the adjoint representation of ${SU_{\rm c}(3)}$. As a class function the characters 
depend on the eigenvalues of $e^{i\theta_i}$ of $g$, (i=1, 2, 3): 
$\chi_{1,0}(g)=\sum_{i=1}^{i=3}e^{i\theta_i}$ with $ \sum_{i=1}^{i=3}{\theta_i}=0$. 
In Eq.~(\ref{eq:Theta}),  $d_G=2$, $d_q=4$ and $d_s=2$ are the number of degrees of
freedom of gluons and quarks, not counting the color multiplicity, 
already taken into account in the characters
since $\chi_{1,1}(0,0)=8$ and $\chi_{1,0}(0,0)=3$.

For high temperature, ($\beta\to 0$), integral~(\ref{eq:ztil}) can be performed by the saddle
point method. The maximum~\cite{tou} of $\Theta$ is reached for $r=0$
 and $\theta_i=0$. Expanding ${\Theta}$ near this point to second
order in $r$ and ${\theta_i}$ one finds
\be
{\Theta}=\frac{1}{3}\frac{v}{\beta^3}u(\mu_q,\mu_s)(1-2 r^2) -
\frac{v}{2\beta^3}C(\mu_q,\mu_s)\sum_{i=0}^{i=3}\theta_i^2 \,.
\label{eq:expa}
\ee  
Finally, integration over $r$ and $\theta_i$ gives in the
saddle point approximation  the physical canonical partition function (after Wick rotation)
\be
{Z}_{Q}(\beta,v,\mu_q,\mu_s)=
\frac{3}{4}\sqrt{2\pi}C^{-4}(\mu_q,\mu_s)\,u^{-\frac{3}{2}}
\left(\frac{\beta^3}{v}\right)^{\frac{13}{2}}
\exp\left[\frac{1}{3}\frac{ u v}{\beta^3}\right] \,,
\label{eq;ZQ}
\ee
where $u(\mu_q,\mu_s)$ are known~\cite{tou} functions of quark 
chemical potentials and
gluon and quark  degrees of freedom (see Eqs.~(\ref{eq:fnu})
and (\ref{eq:fnc}) below for general case of $N$ colors).

\subsection{The bag density of states and BBSBM mass spectrum}\label{subsec:bds}
The density of states $\sigma(W,v,\mu_q,\mu_s)$ of a bag of volume $v$, at energy 
$W$, is derived from its partition function 
$Z_{Q}(\beta,v,\mu_q,\mu_s)$, Eq.~(\ref{eq;ZQ}), through 
inverse Laplace transform:
\be
\sigma(W,v,\mu_q,\mu_s)=\frac{1}{2\pi i}\int_{c+
i\infty}^{c+i\infty}d\beta \exp(\beta W) Z_{Q}(\beta,\mu_q,\mu_s)\,.
\label{eq:bagden}
\ee
For large $W$  this integral can be evaluated 
by the steepest descent method to obtain the asymptotic bag level density:
\be
\sigma(W,v,\mu_q,\mu_s)=\frac{3}{8} C^{-4} u^{7/2}
v^{-3/2} W^{-11/2}\exp\left[4/3\, u^{1/4 }
v^{1/4} W^{3/4}\right]\,.
\label{eq:baglev} 
\ee

 The GCPF of a gas of hadrons (bags) with mass $m_i$ and proper volume $v_i$
 can written in the Boltzmann approximation~\cite{hag3} 
\bea
Z_{H}(T,V,\lambda_q,\lambda_s)&\hspace{-0.0cm}=&\hspace{-0.0cm}\sum^\infty_{N=0}
\frac{1}{N!}\int\prod^N_{i=1}
\left[\frac{d^3p_i}{(2\pi)^3}dm_i\,dv_i\,\left(V-
\sum^N_{j=1}v_j\right)\right.\nonumber\\ &\hspace{+0.2cm}&\hspace{-0cm}\times 
\left. e^{-\beta\sqrt{p^2_i+m^2_i}}\,
\tau(m_i,v_i,\lambda_q,\lambda_s)\vphantom{\sum^N_{j=1}}\right]
\theta\left(V-\sum^N_{j=1}v_j\right)\,,
\label{eq:ZH}
\eea
where $\tau(m,v,\lambda_q,\lambda_s)\,dm\,dv$ is the number of bag states
in the mass interval $(m,m+dm)$ and volume interval $(v,v+dv)$. It has been 
shown~\cite{zino} that the $v$ dependence of $\tau(m,v,\lambda_q,\lambda_s)$ is
 obtained through the derivative 
of $\sigma(W,v,\lambda_q,\lambda_s)$, as given by 
Eq.~(\ref{eq:baglev})  
with respect to the volume $v$,
\be
\tau (m,v,\lambda,\lambda_s)= \left.\frac{\partial}{\partial
v}\sigma(W,v,\lambda,\lambda_s)\right|_{W=m- Bv}\,,
\label{eq:tau}
\ee
where the bag model relation $W=m-Bv$ between the energy 
$W$ and the mass $m$ is used,  $B$ being the bag constant.
This yields the BBSBM asymptotic hadronic mass spectrum:
\bea
\tau (m,v,\lambda,\lambda_s)&\hspace{-0.0cm}\approx 
  &\hspace{-0.0cm}\frac{1}{8}C^{-4} u^{15/4}
v^{-9/4}(m- Bv)^{-19/4}\nonumber\\
&\hspace{-0.5cm}&\hspace{-0.5cm}\hspace{+0.4cm}\times
   \exp{\left[{4\over3}\,u^{1/4}v^{1/4}(m- Bv)^{3/4}\right]}\,.
\label{eq:tauas}
\eea
This is the asymptotic expression of $\tau(m,v,\lambda_q,\lambda_s)$
derived from the asymptotic
expression of $\sigma(W,v,\lambda,\lambda_s)$. For small
masses a phenomenological  function $\tau_0$ is introduced
\be
\tau_0(m,v,\lambda_q,\lambda_s)=\sum_ig_i(\lambda_q,\lambda_s)\,
\delta(m-m_i) \,\delta(v-v_i)\,,
\label{eq:tau0}
\ee
where $g_i(\lambda_q,\lambda_s)$ is the multiplicity modified
by fugacities of a hadron of mass $m_i$ and volume $v_i$.

\section{Description of the ${\rm HG}\ra {\rm QGP}$ phase transition}\label{PT}
\subsection{Pressure partition function}
We study here the analytic properties of the GCPF
to show that the system undergoes a phase transition from  hadron to  QGP phase,
and we implement the generalization to N-colors.  
The analytic properties of the GCPF $Z_{H}(T,V,\lambda_q,\lambda_s)$, Eq.~\ref{eq:ZH}, 
in the thermodynamic limit, ($V\to \infty$), are most easily investigated by
 considering the corresponding pressure-GCPF  which is its Laplace transform with 
respect to the volume:
\be
\Pi(T,s,\lambda_q,\lambda_s)\equiv
\int^\infty_0dV\,e^{-sV}Z_{\rm H}(T,V,\lambda_q,\lambda_s).
\label{eq:Pi}
\ee
Substituting Eq.~(\ref{eq:ZH}) in Eq.~(\ref{eq:Pi}), one obtains
\be
\Pi(T,s,\lambda,\lambda_s)=
{1\over{s-f(\beta,s,\lambda,\lambda_s)}}\,,
\label{eq:PiPi}
\ee
where
\be
f(\beta,s,\lambda_q,\lambda_s)=\!f_0+\!
{1\over2\pi^2\beta}\!\int^\infty_{V_0}\!\!dv
\!\int^\infty_{M_0+Bv}\!\!\!dm
m^2K_2(m\beta)\,\tau(m,v,\cdots)\,e^{-sv}\,,
\label{eq:f0f}
\ee
and where $f_0\equiv f_0(\beta,s,\lambda_q,\lambda_s)$ 
corresponds to the contribution coming from
$\tau_0$. The lower limits $M_0$ and $V_0$ have values above which the
asymptotic expression~(\ref{eq:tauas}) begins to hold.

Because of the properties of the inverse Laplace transform the behavior of 
$Z_{H}(T,V,\lambda_q,\lambda_s)$ in the thermodynamic limit is governed by the rightmost
 singularity of $\Pi(T,s,\lambda_q,\lambda_s)$. From Eq.~(\ref{eq:PiPi}) it can be 
shown~\cite{kat} that $\Pi$  has two  singularities on the real axis of the s-plane: 
a pole $s_0$ given by
\be
s_0(\beta,\lambda_q,\lambda_s)=f(\beta,s_0(\beta,\lambda_q,\lambda_s))\,,
\label{eq:s0}
\ee
and a singularity $s_c(\beta,\lambda_q,\lambda_s)$ 
induced by  $f(\beta,s,\lambda_q,\lambda_s)$ which remains
{\em finite} for $s \to s_c$. Furthermore $s_c$ is found
to be given by
\be
s_c(\beta,\lambda_q,\lambda_s)=\frac{1}{3}u(\lambda_q,\lambda_s)
(\beta^{-3}- B\beta)\,.
\label{eq:sc}
\ee
For given values of $\lambda_q$  and $\lambda_s$ the 
transition temperature
 $T_C(\lambda_q,\lambda_s)=\beta_c^{-1}$ is determined by
matching of these  two singularities, namely when
\be 
s_c(\beta_c(\lambda),\lambda)=s_0(\beta_c(\lambda),\lambda)
=f(\beta_c(\lambda),s_c(\beta_c(\lambda),\lambda),\lambda)\,,
\label{eq:match}
\ee
where we have written $\lambda $ instead
 of $(\lambda_q,\lambda_s)$ to simplify the notation.

\subsection{Properties of hadronic phases}
For $T<T_C$, in the hadron phase, the rightmost singularity
is $s_0(\beta,\lambda_q,\lambda_s)$ and the partition function is governed by this 
singularity in the thermodynamic limit. Then the pressure and
the energy density are given by
\be
\left.\begin{array}{rll}
\displaystyle P_H(T,\lambda)&\displaystyle
\equiv\phantom-\lim_{V\to\infty}\frac{T}{ V}\ln Z_H(T,V,\lambda)
&\displaystyle =\phantom-T\,s_0(\beta,\lambda)\,,\\\\
\displaystyle\varepsilon_H(T,\lambda)&\displaystyle
\equiv-\lim_{V\to\infty}\frac{\partial}{\partial\beta}
\ln Z_H(T,V,\lambda)
&\displaystyle =-\frac{\partial}{\partial\beta}\,s_0(\beta,\lambda)\,.
\end{array}\hspace{1cm}\right\}
\ee 
For $T>T_C(\lambda_q,\lambda_s)$, the leading singularity
is $ s_c$ and we have in the thermodynamic limit
\be
\left.\begin{array}{rll}
\displaystyle P_Q(T,\lambda)&\displaystyle \equiv\phantom-T\,s_c(\beta,\lambda)
&\displaystyle =\frac{1}{3}\,u(\lambda)\,T^4- B\,,\\\\
\displaystyle\varepsilon_Q(T,\lambda)&\displaystyle
\equiv-\frac{\partial}{\partial\beta}s_c(\beta,\lambda)
&\displaystyle = u(\lambda)\,T^4+ B\,.
\end{array}\hspace{1cm}\right\}
\label{eq:plasma}
\ee

\begin{figure}[t]
\vspace*{-0.5cm}
\centerline{\psfig{height=10cm,figure=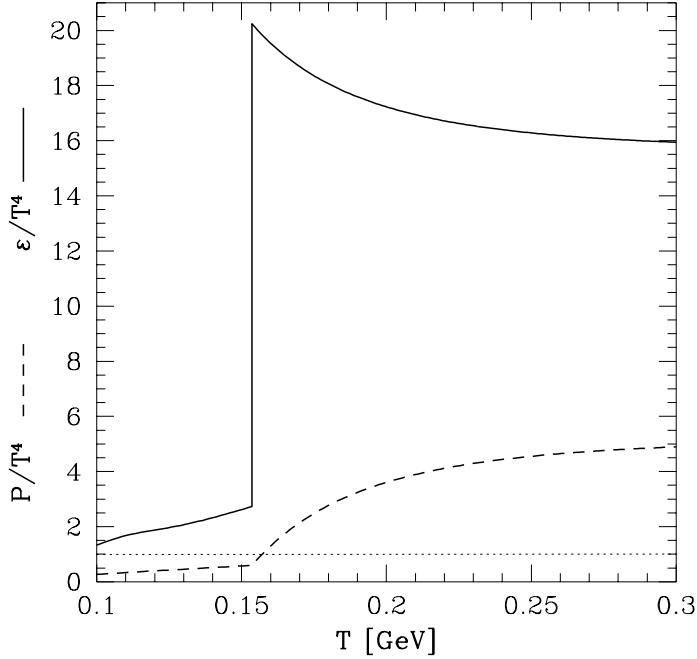 }}
\vspace*{-1cm}
\caption{Energy density  and pressure versus temperature: 
solid curve gives the EoS in the case $\mu_q=\mu_s=0$ and for $B^{1/4}= 225$
 MeV corresponding to $T=150$ MeV. Dashed line is the pressure in the
same conditions.\label{fig:eps }}
\end{figure}

\begin{figure}[bht]
\vspace*{-0.5cm}
\centerline{\psfig{height=10cm,figure=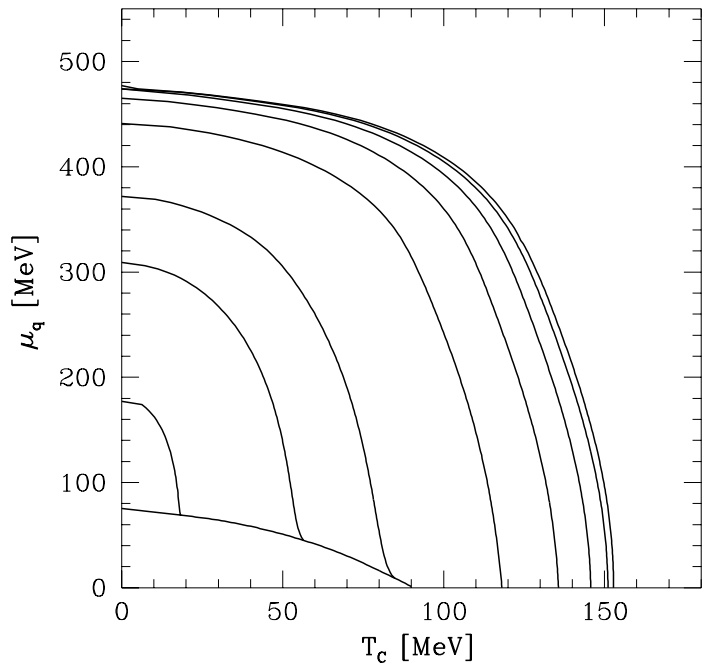 }}
\vspace*{-0.8cm}
\caption{Critical curves in the plane $T$--$\mu_q$ for
$\mu_s=$ 0, 100, 200, 300, 400, 500, 536, 560 MeV, right to left. The lower curve
corresponds to $\bar K$ condensation. \label{fig:dphse}}
\end{figure}

Eqs.~(\ref{eq:plasma}) are the equations of a non interacting QGP. 
Of course other quantities like baryon number density, quark number 
densities can be derived form the expression of the partition function in the 
thermodynamic limit.
Fig.~\ref{fig:eps } displays the energy density and the pressure
 for both phases
for a value of  $B^{1/4}=225$ MeV, corresponding 
to  $T_C=150$ MeV. Fig.~\ref{fig:dphse}
is the phase diagram obtained by solving Eq.~(\ref{eq:match}) for different values 
of the strange quark chemical potential $\mu_s$.


\subsection{Extension to $N$-colors}
The previous results can naturally be extended~\cite{toun1} to 
$SU_{\rm c}(N)$. By the same  method as in 
Sec.~\ref{subsec:bpf} and Sec.~\ref{subsec:bds},
 for  $N=3$,
we obtain for the grand canonical partition function
\be
{ Z}_{Q}(\beta,v,\mu_q, \mu_s)=A\,C^{(1-N^2)/2}u^{-3/2}
{(\frac{v}{\beta^3})}^{(N^2+4)/2}\exp\left[\frac{1}{3}
\frac{u v}{\beta^3}\right]\,,
\label{pfn}
\ee
where
\be
A=\displaystyle{\frac{3\sqrt{3}\prod_{i=1}^{N-1}i\,!}{8\,{(2\pi)}^{(N-4)/2}\,
\sqrt{N}}}\,,
\ee
\bea
u(\mu_q,\mu_s)&=&\frac{\pi^2}{30}\left[(N^2-1)\ d_g+
\frac{7}{4}N(d_q+d_s)+\frac{15\,N}{2}d_q
\left(\frac{\beta\mu_q}{\pi}\right)^2\! 
\left(1+\frac{\beta^2\mu_q^2}{2\pi^2}\right)\right.\nonumber \\
&\hspace{-0.3cm}&\hspace{-0.3cm}\hspace{2.5cm}\left.
+\frac{15}{2}N\,d_s\left(\frac{\beta\mu_s}{\pi}\right)^2\! 
\left(1+\frac{\beta^2\mu_s^2}{2\pi^2}\right)\:\right] \,,
\label{eq:fnu}
\eea
and
\be
C(\mu_q,\mu_s)=\left[\frac{N}{3}\,d_g +\frac{d_q+ds}{6}+
\frac{d_q}{2}\left(\frac{\beta\mu_q}{\pi}\right)^2+
\frac{d_s}{2}\left(\frac{\beta\mu_s}{\pi}\right)^2\right] \,.
\label{eq:fnc}
\ee

By inverse Laplace transform of $Z_{Q}(\beta,v,\mu_q, \mu_s)$, 
we obtain the bag density of levels  $\sigma(W,v,\mu_q,\mu_s)$ for large $W=m-Bv$,   
from which the hadronic mass spectrum $\tau(m,v,\mu_q,\mu_s)$ for large
$m$ is derived, according to Eq.~(\ref{eq:tau}) 
\bea
\tau(m,v,\mu_q,\mu_s)\sim C^{(1-N^2)/2}u^{3(1+N^2)/8}v^{\gamma}\,W^{\delta}
\exp\left[4/3\, u^{1/4 }v^{1/4} W^{3/4}\right]\,, 
\label{eq:tauN}
\eea
where $\gamma=-(N^2+9)/8$ and $\delta=-(11+3N^2)/8$.
 Furthermore it can be shown~\cite{kat,toun1} that the conditions for a phase 
transition to occur are $\gamma +\delta=-(5+N^2)/2 <-3$ and
$\delta=-(11+3N^2)/8 <-7/4$. Both conditions give $N^2 > 1$. This can also be
 seen on the mass dependence of the spectrum. From the bag relations, $m=4Bv=W+Bv$, 
we have $v= m/4B$ and $W=3m/4$. Substituting for $W$ and $v$ in Eq.~(\ref{eq:tauN}) we find 
\be
\tau(m,\mu_q,\mu_s)\sim\,m^{-(N^2+5)/2}C^{(1-N^2)/2}
u^{3(1+N^2)/8}B^{(9+N^2)/8}  
\exp{\left[\left(\frac{u}{3\,B}\right)^{1/4}\!m\right]}\,,
\label{eq:taum}
\ee
where $u$ and $C$ are given by Eq.~(\ref{eq:fnu}) and  Eq.~(\ref{eq:fnc}). 
Comparison with a spectrum of the form in
 Eq.~(\ref{eq:sol})
shows that $a =-(N^2+5)/2$ which is, for $N>1$,  smaller than $-7/2$
as required for a phase transition to occur (see Sec.~\ref{subsec:partstb}). 
Thus for $N=1$ there is no phase transition. For $N=2$  our approach is not
equivalent to lattice-QCD where one find a second order phase transition~\cite{laer}, 
a result requiring further study.

\section{Conclusions}
Needless to repeat here, lattice QCD and
its continuum limit is the fundamental theory for describing
in a unified scheme the deconfinement phase transition, 
also for {\em finite} baryon density,
and we are looking forward to future developments in this direction. 
As we await these advances, we have offered here a relatively simple 
extension of the statistical bootstrap model, motivated by some well
understood properties of strong interactions. We have shown that in 
theoretically consistent manner we can study the properties of 
deconfinement phase transition at finite baryon density. 
There are several simplifications in our approach which need attention:
for example we assumed vanishing quark mass even for the s-quark, and  we
ignored QCD interactions in the bag. Even so, 
 we find several interesting results:  the properties of the 
latent heat of the phase transition, and the general features of the 
hadronic mass spectrum are in as one could wish for, 
and there is no  phase transition when $N=1$.

%

\end{document}